\documentclass{iau379}

\usepackage{graphicx}
\usepackage{amsmath}
\usepackage{multirow}

\newcommand{\Gaia}{{\it Gaia}\,}

\begin{document}

\lefttitle{Someone et al.}
\righttitle{IAU Symposium 379: Template}

\jnlPage{1}{7}
\jnlDoiYr{2023}
\doival{10.1017/xxxxx}

\aopheadtitle{Proceedings of IAU Symposium 379}
\editors{P. Bonifacio,  M.-R. Cioni, F. Hammer, M. Pawlowski, and S. Taibi, eds.}

\title{The Formation of Magellanic System and the total mass of Large Magellanic Cloud}

\author{Jianling WANG$^1$, Francois Hammer$^2$, Yanbin Yang$^2$, Maria-Rosa L. Cioni$^3$, }
\affiliation{$^1$ CAS Key Laboratory of Optical Astronomy, National Astronomical Observatories, Beijing 100101, China\\
$^2$ GEPI, Observatoire de Paris, CNRS, Place Jules Janssen, F-92195 Meudon, France \\
$^3$ Leibniz-Instit\"{u}t f\"{u}r Astrophysik Potsdam, An der Sternwarte 16, D-14482 Potsdam, Germany
}

\begin{abstract}
The Magellanic Stream is unique to sample the MW potential from $\sim$50 kpc to 300
kpc, and is also unique in constraining the LMC mass, an increasingly important
question for the Local Group/Milky Way modeling.  Here we compare 
strengths and weaknesses of the two types of models (tidal and ram-pressure) of
the Magellanic Stream formation. I will present our modeling for the formation
of the Magellanic System, including those of the most recent discoveries in the
Stream, in the Bridge and at the outskirts of Magellanic Clouds. This model has
been successful in predicting most recent observations in both properties of
stellar and gas phase. It appears that it is an over-constrained model and
provides a good path to investigate the Stream properties. In particular, this
model requires a LMC mass significantly smaller than 10$^{11}$ M$_\odot$.
\end{abstract}


\maketitle

\section{Introduction}

The Magellanic Stream (MS) and Leading Arm (LA) subtend an angle of
230$^{\circ}$,  which is identified to be anchored to the Magellanic Clouds in
1974 by \citet{Mathewson1974}. The nature of its formation was
considered still unknown in 2012 \citep{Mathewson2012}.  Modern observations of
proper motion from both HST and GAIA are indicating that the the Clouds are
presently at first passage to the Milky Way \citep{Kallivayalil2006,Piatek2008,Kallivayalil2013}. 
Besides large amount of neutral gas distributed
along the Stream, there are mounting evidence that 3-4 times more ionized than neutral
gas has been deposited along the Stream \citep{Fox2014,Richter2017}. 

In the first infall frame, the explanations of the MS can be broadly classified
into two schemes. One is the tidal tail model \citep{Besla2012,Lucchini2020}, 
the other is ram-pressure tails \citep{Mastropietro2010,Hammer2015,Wang2019,Wang2022}.  
In the tidal tail model, the MS is
generated by the mutual close interaction 1-2 Gyr ago before the MCs entering
into the halo of MW. In this scenario, the SMC is assumed to be a long-lived
satellite of LMC, which requires a LMC mass in excess of 10$^{11}$ M$_{\odot}$.

There are several major limitations for the tidal model. First, it already
lacks by a factor of 10 the amount of neutral gas observed in the Stream.  Second, it is unable
 to reproduce the huge amount of ionized gas that is observed along the Stream.
Third, it can only produce a single stream filament, while the MS are made of two
filaments, which have been clearly identified by chemical, kinematic, and
morphological analyses \citep{Nidever2010,Hammer2015}. Fourth, no
stars along the stream have been observed. Recent revised tidal model have been
made by including hot corona of LMC to amend parts of above drawbacks
\citep{Lucchini2020,Lucchini2021}. But these models
required either a unreasonable massive corona of LMC that is even larger than
that of MW, or a dramatic change of the Cloud orbits, which can not
reproduce their observed proper motions within $3\sigma$.

Conversely to the tidal model, the ram-pressure plus collision model
\citep{Hammer2015,Wang2019} naturally reproduce most observational properties
associated with the Magellanic System, for instance, the dual filaments, huge
amount of ionized gas, absence of stars in the Stream. Interestingly, several observations made after the elaboration of the model have been reproduced without fine tuning.

%With the progress of future observations, more and more predictions from this model have been confirmed. 

In this scenario, the Leading Arm are the trailing gas of front-runner dwarfs
\citep{Yang2014,Hammer2015,Tepper-Garcia2019}, which is well supported by the determination of low metal abundances in 3 part of this structure (see Philip Richter's contribution).

\section{Two hydrodynamic filaments formed by ram-pressure plus collision}

\begin{figure}
  \includegraphics[scale=.37]{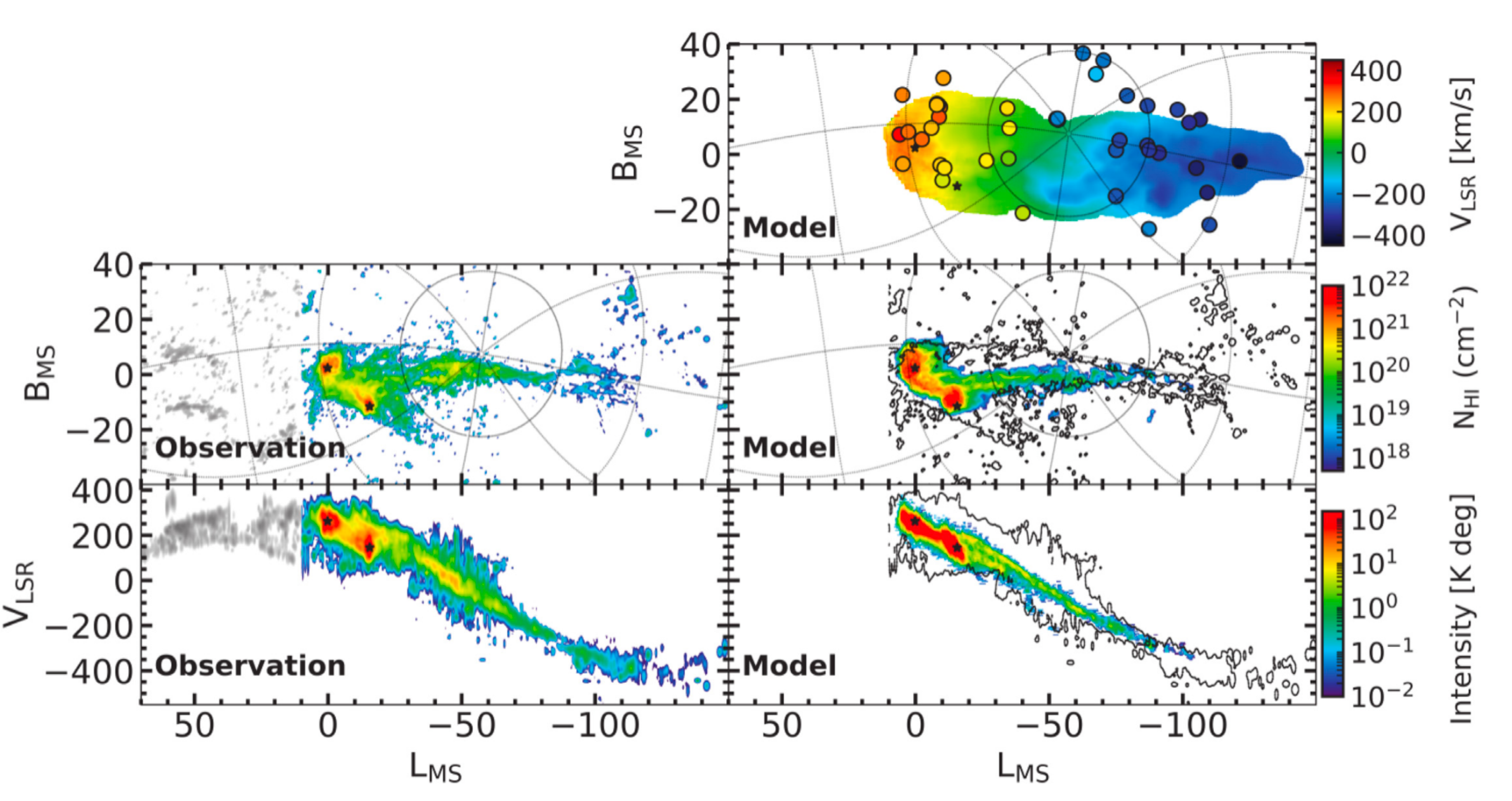}
  \caption{Comparing HI and ionized gas from simulation model (right panels)
from \citet{Wang2019} with observations (left panels) of Nidever et al.
(2010). The top-right panel shows the sky distribution of the simulated ionized
gas with a color coding for the line-of-sight velocity. Circle points
represent QSOs absorption line observations by HST/COS \citep{Fox2014}. The
simulated ionized gas mass is consistent with that observed \citep{Donghia2016}. The bottom two rows compare observed HI distributions of the Magellanic
Stream with that of simulations. The black stars in each panel indicate the
position of LMC and SMC. In the simulation panels the contours indicates the
observations data.  The LA is assumed to have another origin rather than the MC
gas \citep{Yang2014,Hammer2015,Tepper-Garcia2019}.}
  \label{MS_HI}
\end{figure}

\begin{figure}
\center
  \includegraphics[scale=.37]{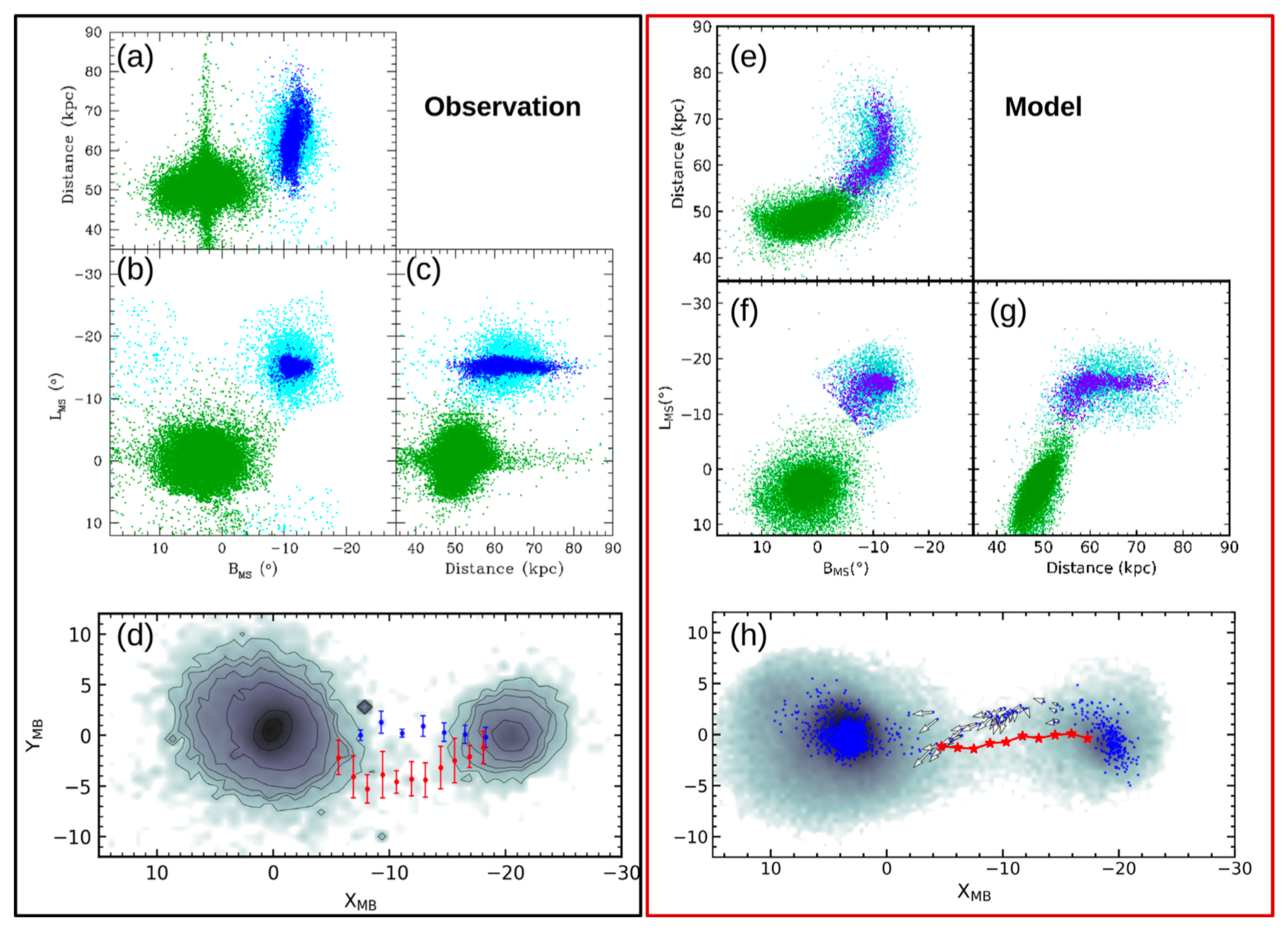}
  \caption{Comparing the stellar distribution of MCs between simulation model
from \citet{Wang2019} (the right-hand panels enclosed within rectangle a red
box ) and observation data (the right-hand panels enclosed within a black box).
In the observation panels (a,b,c), the green and cyan points represent ancient
RR Lyrae stars from LMC and SMC, and blue points indicates the Classical
Cepheids \citep{Ripepi2017}. In the simulation panels, particle numbers and
sky distribution have been selected following the observations. The simulation 
model reproduces well the 'cigar' shape of SMC. Panel d shows the ancient star 
distribution on the sky, overlapping in the Bridge region RR Lyrae stars and 
young main sequence stars from \citet{Belokurov2017}. Panel h shows simulation 
model from \citet{Wang2019} which show a similar offset of young stars with ancient 
stars in the Bridge region.} 
\label{SMC} 
\end{figure}

In the frame of the ram-pressure and collision model, we have built a stable model of
Milky Way which include a hot gas corona. The progenitors of MCs are gas rich
dwarf galaxies before entering the halo of MW. Figure \ref{MS_HI} compare the
observed neutral gas and ionized gas to this model. This model naturally
generates two HI streams behind the MCs, and a huge amount of ionized gas
deposited along the stream. The strong mutual interaction between the MCs
totally stretched by gravitational tides the SMC into a 'cigar' shape, which is well reproduced by this
model as shown in Figure \ref{SMC}. Recent observation indicates that there is
offset between the ancient stars and young stellar population in the Bridge
region \citep{Belokurov2017}, which is also well reproduced in this model.

\section{Many Predictions Are Confirmed by Observations}

With the progress of observations, new data and findings provide essential test 
for any model aiming at reproducing the formation of Magellanic System. We will show that 
our ram-pressure plus collision model pass these tests with many predictions 
confirmed by recent observations. 

\subsection{Two separated populations in the Bridge region} 

\begin{figure}
\includegraphics[scale=.40]{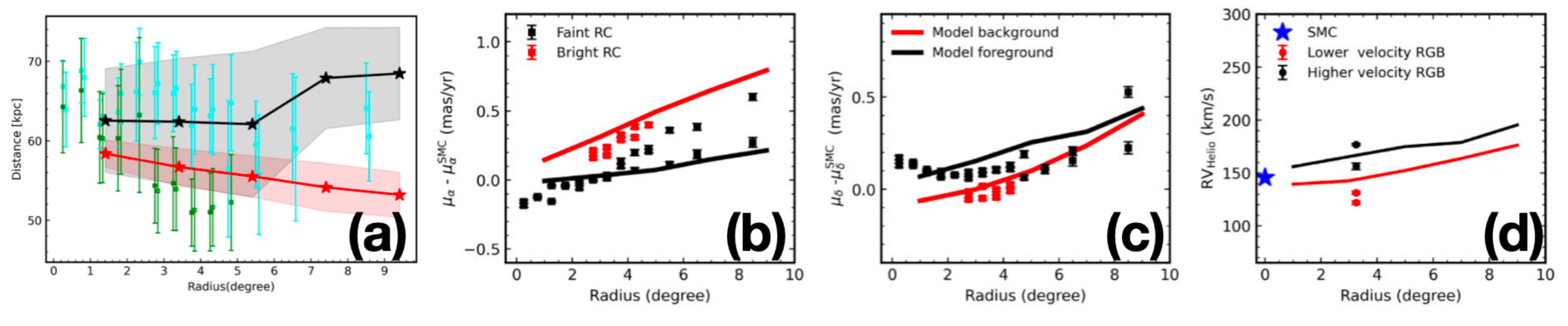} \caption{Comparing the observed
distances (panel a), relative proper motion (panel b, c), and line-of-sight
velocity (panel d) of foreground and background stellar population in the
Bridge region as function radius to SMC between observation data and simulation
model. In panel a, the green and cyan color points indicate bright and faint
population of red clump stars from \citet{Omkumar2021} for north-east (solid
square) and south-east region (open square). The proper motion of faint (black
square) and bright (red square) RC in the left-hand and middle panel are from
\citet{Omkumar2021}. The radial velocity of the lower (red circles) and
higher (black circles) velocity RGB stars are from \citet{James2021}, which
are corresponds to foreground components and main body of SMC. The blue star 
in the panel d indicates the SMC value. The observed proper motions and radial
velocity of SMC are from \citet{Zivick2018}.} 
\label{bridge} 
\end{figure}

Observations indicate that there are two different populations in the
Bridge region, which are separated in both distance and kinematics space
\citep[e.g.,][]{Omkumar2021,James2021}. \citet{Omkumar2021} found that two
populations of red clump stars in the Bridge region starting from SMC to LMC,
which show different brightnesses. The bright and faint red clump populations
show different distance and kinematics, which consistent with the finding of
\citet{James2021}. In our model, the two populations are formed by SMC, which
are tidally stripped by the LMC. The foreground population indicates the disk
component of SMC, which is tidally stripped from SMC and showing debris of
interaction stretching from SMC to LMC. While the background population come
from the spheroid component which is less affected by the LMC tidal and
distributed in the back of the Bridge region.  This model naturally reproduces
this new observation as shown in Figure \ref{bridge}.

\subsection{Periphery of the Clouds}

\begin{figure}
  \center
  \includegraphics[scale=.45]{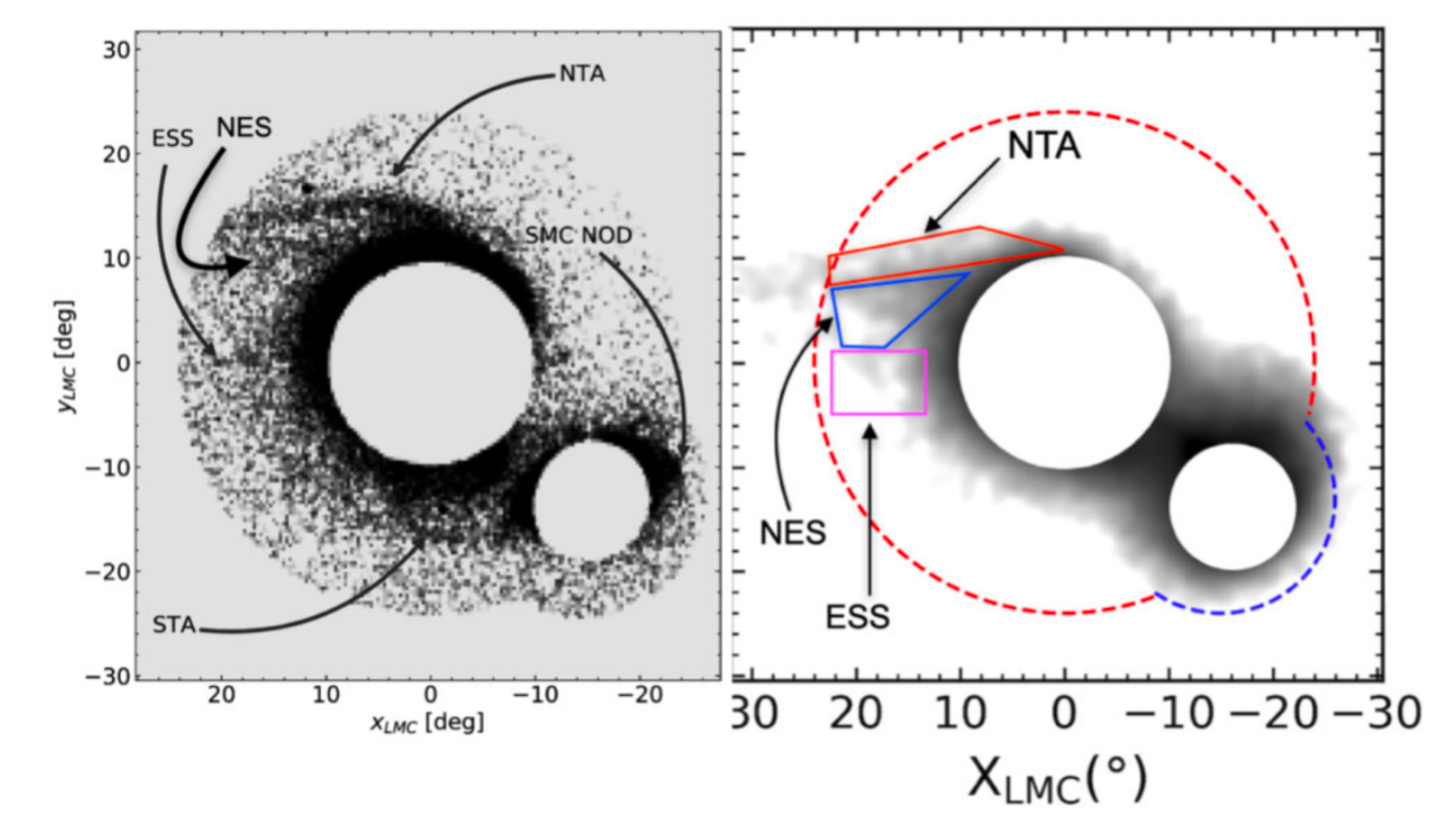}
  \caption{Comparing the morphology of MCs with Gaia EDR3 data \citep{Luri2020} 
 with simulation model \citep{Wang2022}.  In the right
panel, the red and blue dashed line indicate the sample selection region for
Gaia data \citep{Luri2020}. In the right-hand panel, the colored
polygon regions indicate different substructures associated with LMC detected
by \citet{Gatto2022}.}
  \label{NTA}
\end{figure}

\begin{figure}
  \center
\includegraphics[scale=0.32]{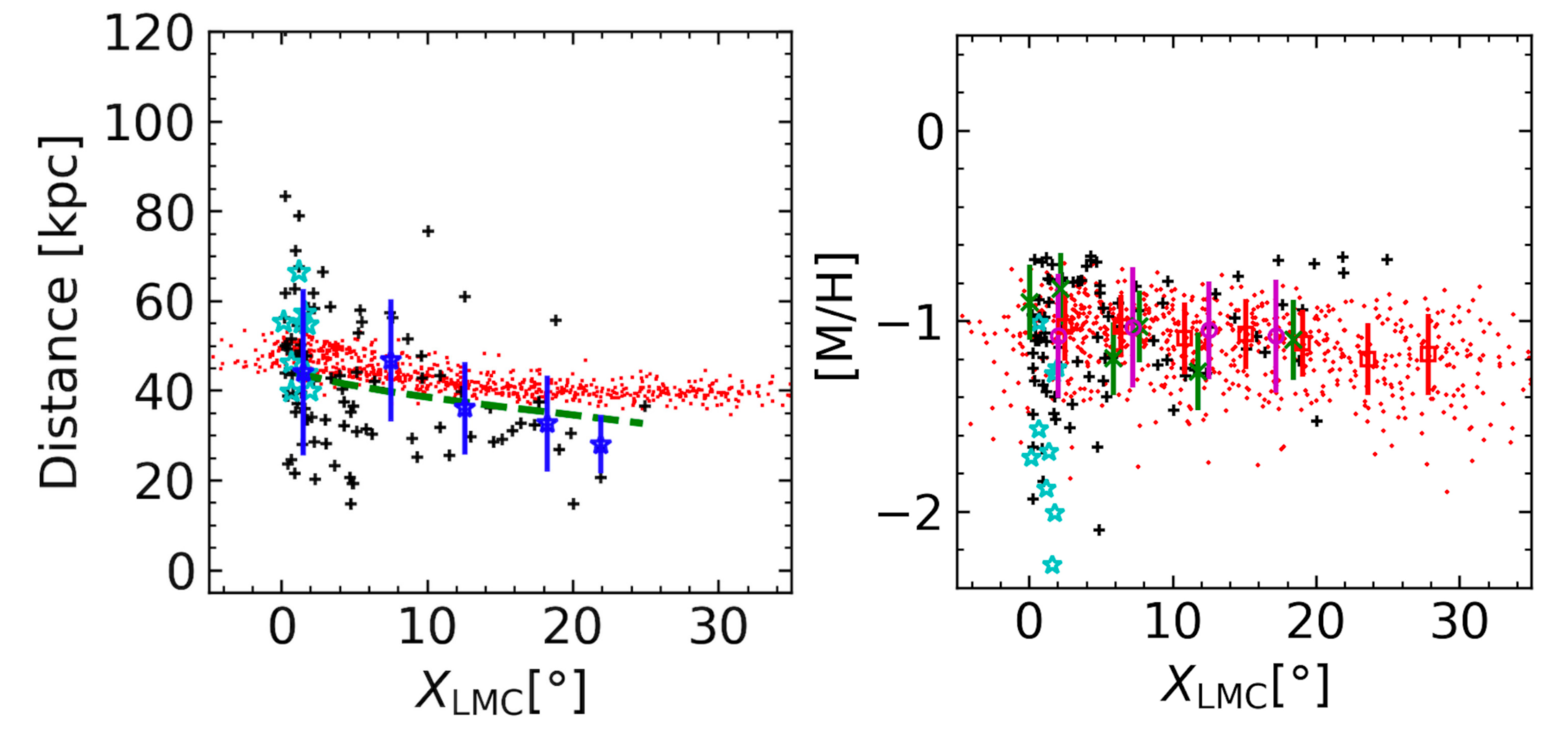}
\caption{The distance and metallicity of NTA stars varied with
X$_\mathrm{LMC}$. The black crosses are selected with Gaia EDR3 \citep{Luri2020} and StarHorse \citep{Anders2022} 
(see the text for details), and the red points indicates data from our simulation model. The cyan
stars indicate the data from \citet{Huang2022}, who derived star distance and
metallicity from SkyMapper DR2 \citep{Onken2019} and Gaia EDR3. The blue
stars and its associated error bars in the left panel indicate the mean and
dispersion for the observed stars. The green line in the left panel denotes an
inclined disk following \citet{vdM2014}. In the right
panel, the magenta squares and its associated error bars show the mean and
dispersion of observation data from \citet{Grady2021}. The green
stars and its associated error bars indicate the value from observation data of
MagES \citep{Cullinane2022a}. The red squares and its associated error bars
are the mean and dispersion value of simulation model.}
\label{NTAdz}
\end{figure}

%\begin{figure}
%  \includegraphics[scale=.37]{Periphery_Kine.pdf}
%  \caption{An example figure with space for artwork.}
%  \label{kine}
%\end{figure}

With deep observations, many faint features in the periphery of MCs have been
discovered as shown in the left panel of Figure \ref{NTA} from
\citet{Luri2020}, many of which have been well predicted by our model as shown
in the right panel of Figure \ref{NTA}. The North Tidal Arm (NTA) is the
largest tidal features rooted from the disk LMC, which is confirmed to originate from the LMC on the basis of its metallicity, distance, and
kinematics \citep{Cullinane2022a,Cullinane2022b,Cullinane2020}. Before observations, the model of
\citet{Wang2019} predicted the existence of the NTA , which is formed 
by the Galactic tides exerted on the disk of LMC.

From Gaia EDR3, we have selected stars belonging to NTA according to its
morphology position, proper motion. With this sample stars of NTA, we
cross-matched with results of literature to get their distance and metallicity.
The distance and metallicity variation as function of radius to the LMC are
shown in Figure \ref{NTAdz}. 

In order to compare simulation model with observations,  we assigned
metallicities to particles of the simulation model by painting that of the
LMC with metallicities following the observational constraints \citep{Grady2021}.
\citet{Grady2021} selected red giant stars of LMC from \Gaia DR2, and used
machine-learning method with data of Gaia+2MASS+WISE to estimate the
photometric metallicity for these stars. With these data set, they estimated
radial metallicity profile: [Fe/H] = $\alpha$ R + $b$.  They found the
$\alpha=-0.048\pm0.001$ dex kpc$^{-1}$ and $b=-0.656\pm0.004$ dex. We use this
relation to paint the initial metallicity of our modeled LMC.  For simplicity, we
have adopted the approximated relation [Fe/H] $\sim$ [M/H]. In Figure
\ref{NTAdz}, the modeled data are shown with red points. The simulation model, can explain well the observational results for both
the distance and metallicity profiles, without fine tuning. The current observed metallicity profile 
of the LMC \citep{Grady2021} also reproduces the formation of NTA, which indicates 
that the LMC metallicity profile has been settled down before the formation of NTA, 
or the mutual interaction of MCs/gas loss have marginal influence on the metallicity 
structure of LMC.

\section{Conclusion}

The ram-pressure plus collision model \citep{Hammer2015,Wang2019} can not only
reproduce MS, but also succeed in predicting many observations that have been done in the meantime (see a description in \citealt{Wang2022}).
This model naturally reproduces the two inter-twisted filaments of HI MS, as well
as the huge amount of ionized gas associated with MS. This ability 
also validates that this model goes into the right direction to disentangle the
mystery of Magellanic System formation (Mathewson, private communication).  We
conjecture that the LMC mass has to be small (a few times  10$^{10}$
M$_{\odot}$) to form the Magellanic Stream, though further studies are needed
to explore the exact mass range.

\begin{discussion}

\discuss{Müller Oliver}{Do you think we can find other dwarfs showing 'cigar' shape as SMC in the extragalactic dwarfs ?}
\discuss{Jianling WANG}{This is difficult, since we need precise distance for individual star showing dwarf shapes in 3D. }

\end{discussion}

\end{document}